\documentclass[12pt]{article}
\newcommand{\beq}{\begin{eqnarray}}
\newcommand{\eeq}{\end{eqnarray}}
\newcommand{\drawsquare}[2]{\hbox{%
\rule{#2pt}{#1pt}\hskip-#2pt
\rule{#1pt}{#2pt}\hskip-#1pt
\rule[#1pt]{#1pt}{#2pt}}\rule[#1pt]{#2pt}{#2pt}\hskip-#2pt
\rule{#2pt}{#1pt}}

\newcommand{\Yfund}{\raisebox{-.5pt}{\drawsquare{6.5}{0.4}}}
\newcommand{\Ysymm}{\raisebox{-.5pt}{\drawsquare{6.5}{0.4}}\hskip-0.4pt%
        \raisebox{-.5pt}{\drawsquare{6.5}{0.4}}}
\newcommand{\Yasymm}{\raisebox{-3.5pt}{\drawsquare{6.5}{0.4}}\hskip-6.9pt%
        \raisebox{3pt}{\drawsquare{6.5}{0.4}}}

\newcommand{\jref}[4]{{\it #1} {\bf #2}, #3 (#4)}

\newcommand{\NC}[3]{\jref{Nuovo Cim.}{#1}{#2}{#3}}
\newcommand{\NPB}[3]{\jref{Nucl.\ Phys.}{B#1}{#2}{#3}}

\newcommand{\PLB}[3]{\jref{Phys.\ Lett.}{#1B}{#2}{#3}}

\newcommand{\PRD}[3]{\jref{Phys.\ Rev.}{D#1}{#2}{#3}}

\newcommand{\PRL}[3]{\jref{Phys.\ Rev.\ Lett.}{#1}{#2}{#3}}

\makeatletter
\def\vereq#1#2{\lower3pt\vbox{\baselineskip1.5pt \lineskip1.5pt
\ialign{$\m@th#1\hfill##\hfil$\crcr#2\crcr\sim\crcr}}}
\makeatother

\begin{document}

\begin{titlepage}
\begin{center}
\today     \hfill    LBNL-42332 \\
~{} \hfill UCB-PTH-98/47  \\
~{} \hfill hep-th/9810014\\

\vskip .3in

{\Large \bf New Confining $N=1$ Supersymmetric Gauge Theories\footnote{This 
work was supported in part by the U.S. 
Department of Energy under Contract DE-AC03-76SF00098, and in part by the 
National Science Foundation under grant PHY-95-14797.}}

\vskip 0.3in

{\bf Csaba Cs\'aki\footnote{Research fellow, Miller Institute for 
Basic Research in Science.} and Hitoshi Murayama\footnote{Supported in 
part by an Alfred P. Sloan Foundation Fellowship.}}

\vskip 0.15in

{\em Theoretical Physics Group\\
     Ernest Orlando Lawrence Berkeley National Laboratory\\
     University of California, Berkeley, California 94720}

\vskip 0.1in
{\rm and}
\vskip 0.1in

{\em Department of Physics\\
     University of California, Berkeley, California 94720}

\vskip 0.1in
{\tt  csaki@thwk5.lbl.gov, murayama@lbl.gov}

\end{center}

\vskip .25in

\begin{abstract}
We examine $N=1$ supersymmetric gauge theories which confine in the presence 
of a tree-level superpotential. We show the confining spectra which satisfy
the 't Hooft anomaly matching conditions and give a simple method to find
the confining superpotential. Using this method we fix the confining 
superpotentials in the simplest cases, and show how these superpotentials
are generated by multi-instanton effects in the dual theory. 
These new type of confining theories may be useful for model building, since
the size of the matter content is not restricted by an index constraint.
Therefore, one expects that a large variety of new confining spectra
can be obtained using such models.
\end{abstract}

\end{titlepage}

\newpage

\section{Introduction}
\setcounter{equation}{0}
\setcounter{footnote}{0}

Confining theories are the simplest asymptotically free $N=1$ supersymmetric 
gauge theories. In such theories the low-energy effective theory is simply 
given by a Wess-Zumino model for the composite gauge singlets. The first
example of such a confining theory has been found by Seiberg~\cite{Seiberg}.
Later several other confining theories have been 
found~\cite{IntPoul,SUanti,SO,Spanti,s-conf,QMM,DM} and classified
(for a recent review see~\cite{Minneapolis}). Some of these confining theories
have been used for constructing models which explain the flavor hierarchy
\cite{flavor}, but the limited number of known confining models 
strongly limits their applications for model building. 
All of the confining theories mentioned above  
are based on examples with simple gauge groups
which confine without the presence of a tree-level 
superpotential. However, it has been noted by Kutasov, Schwimmer and Seiberg in
Ref.~\cite{KSS} that certain theories might be confining if a suitable 
tree-level
superpotential is added to the theory. In \cite{KSS}, the dualities of 
$SU(N)$ with an adjoint field $X$ and $F$ flavors were examined in the 
presence of a tree-level superpotential ${\rm Tr} X^{k+1}$.  They
noted that for a certain number of colors 
($N=kF-1$) the dual gauge group reduces to $SU(1)$, and identified a set 
of composites which satisfy the 't Hooft anomaly matching conditions.

In this paper, we show that the theory is indeed confining
for all values of $k$. These theories can be thought of as generalizations of
the well-known s-confining theories~\cite{s-conf}, since in the case when 
the tree-level superpotential reduces to a mass term ($k=1$ in the above 
example) one always obtains an s-confining theory. However, contrary to the
ordinary s-confining theories, the matter content of these examples
is not restricted by an index constraint. Therefore, we expect that there are 
many more confining theories of this sort exhibiting a large variety of
global symmetries and confining spectra, some of which may be 
useful for composite model building.

In this paper we will give a method to find the
confining superpotential of these theories, which reproduces the classical
constraints once the $F$-flatness conditions arising from the tree-level 
superpotential ${\rm Tr} X^{k+1}$ is taken into account. Once this 
superpotential is established, one can integrate out flavors in order to
find dynamically generated Affleck-Dine-Seiberg-type \cite{ADS} (ADS) 
superpotentials.
However, the confining superpotentials obtained in this paper are not the 
most general ones, since the form of the tree-level superpotential is
assumed to be ${\rm Tr} X^{k+1}$, and the relations resulting from the
requirement of the vanishing of the $F$-terms are used in constructing the
superpotential. Therefore superpotential perturbations along directions 
other than mass terms for some flavors will not be correctly reproduced 
by the superpotentials presented here. In order to reproduce such 
perturbations as well, one would need to find the confining superpotential
in the presence of the most general tree-level superpotential, which we 
leave for future investigation.

In all solutions presented in this paper, we find that the confining 
superpotentials or the ADS superpotentials are always due to multi-instanton 
effects, and not due to a one-instanton effect. As expected, 
the coupling of the 
tree-level superpotential also appears in the dynamically generated 
superpotentials, which diverge in the limit when this coupling is 
turned off.

This paper is organized as follows. In Section~\ref{sec:SU} we first discuss
the confining theories based on $SU(N)$ with an adjoint and fundamentals.
We examine the $k=2$ case in detail, show how to find the confining 
superpotential and how it arises from a two-instanton effect in the 
dual theory. Then we consider integrating out a flavor from the $k=2$
theory. Next we examine the $k>2$ theories. We show that there
are no additional branches in this theory, and thus just as in the 
$k=2$ case they are confining at the origin. We show the confining 
spectrum and write down the form of the confining superpotential (without 
fixing the coefficients of the individual terms).
In Section~\ref{sec:other} we consider generalizations of the 
theories presented in Section~\ref{sec:SU} to theories with more 
complicated gauge group and/or matter content. We find the confining spectrum
for these theories, but in most examples leave the determination of the 
superpotentials for future work. We conclude in Section~\ref{sec:concl}.

\section{The $SU(N)$ theory with an adjoint and fundamentals\label{sec:SU}}
\setcounter{equation}{0}
\setcounter{footnote}{0}

\subsection{The $k=2$ Theories}

Consider $SU(N)$ with an adjoint and $F$ flavors, and a superpotential
${\rm Tr} X^3$ for the adjoint. The global symmetries of the theory are given 
by
\begin{equation}
\begin{array}{c|c|ccccc}
& SU(N) & SU(F) & SU(F) & U(1) & U(1)_R & Z_{3F}\\ \hline
X & Adj & 1 & 1 & 0 & 2/3 & F\\
Q & \Yfund & \Yfund & 1 & 1 & 1-\frac{2N}{3F} & -N\\
\bar{Q} &\overline{\Yfund} & 1 & \Yfund & -1 &  1-\frac{2N}{3F} & -N\\
\end{array}
\end{equation}
This is the special case of the theories considered in~\cite{KSS,K,KS}
for the theory with an adjoint and the superpotential $$W = h {\rm Tr} 
X^{k+1}$$
for $k=2$.  Here, $h$ is a coupling constant, dimensionless for $k=2$, and of 
dimension $k-2$ in general.
For $2F-N>1$ it has been shown in~\cite{KSS,K,KS} that 
the theory has a dual description in terms of the gauge group
$SU(2F-N)$, an adjoint $Y$, dual quarks $q,\bar{q}$ and mesons
$M_1$, $M_2$. The field content and superpotential of the dual theory are 
summarized below:

\begin{equation}
\begin{array}{c|c|ccccc}
& SU(2F-N) & SU(F) & SU(F) & U(1) & U(1)_R & Z_{3F}\\ \hline
Y & Adj & 1 & 1 & 0 & 2/3 & F\\
q & \Yfund & \overline{\Yfund} & 1 & 1 &  1-\frac{2(2F-N)}{3F} & N+F\\
\bar{q} & \overline{\Yfund} & 1 & \overline{\Yfund} & -1 
&1-\frac{2(2F-N)}{3F} & N+F\\
M_1 & 1 & \Yfund & \Yfund & 0 & 2-\frac{4N}{3F} & -2N\\
M_2 & 1 & \Yfund & \Yfund & 0 & \frac{8}{3} -\frac{4N}{3F} & F-2N\\
\end{array}
\end{equation}
\begin{equation} W_{magn}= - h {\rm Tr} Y^3 + \frac{h}{\mu^2}\left( 
M_1 \bar{q}Yq +M_2 \bar{q}q \right). \end{equation}  The coefficients of the 
superpotential terms were fixed to be $-h$ and $h$ by to the analysis 
in~\cite{KSS}.
However, for the case $2F-N=1$ the theory is no longer in the non-abelian 
Coulomb phase (or the free magnetic phase), 
but rather confining. The spectrum can be obtained by adding a 
superpotential term $M_1$ to the 
magnetic $SU(3)$ theory of $2F-N=3$ (which corresponds to integrating out
a single flavor from the electric theory).
The confining spectrum is given by:
\begin{eqnarray}
\label{spectrum}
\begin{array}{c|c|ccccc}
& SU(2F-1) & SU(F) & SU(F) & U(1) & U(1)_R & Z_{3F}\\ \hline
X & Adj & 1 & 1 & 0 & 2/3 & F \\
Q & \Yfund & \Yfund & 1 & 1 &  1-\frac{2(2F-1)}{3F} & 1+F\\
\bar{Q} & \overline{\Yfund} & 1 & \Yfund & -1 &  1-\frac{2(2F-1)}{3F}
 & 1+F\\ \hline \hline 
M_1=(\bar{Q}Q) &  & \Yfund & \Yfund & 0 & 2-\frac{4(2F-1)}{3F} & 
2-F\\
M_2=(\bar{Q}XQ) &  & \Yfund & \Yfund & 0 & \frac{8}{3}-\frac{4(2F-1)}{3F}&
2 \\
B=(Q^F(XQ)^{F-1}) &  & \overline{\Yfund} & 1 & 2F-1 &  1-\frac{2}{3F}&
-1\\
\bar{B} =(\bar{Q}^F(X\bar{Q})^{F-1}) &  & 1 & \overline{\Yfund} & -2F+1 
& 1-\frac{2}{3F} & -1 \\
\end{array} \nonumber \\
\end{eqnarray}
This anomaly matching for the continuous global symmetries $SU(F)^3$, 
$SU(F)^2U(1)$, \break $SU(F)^2U(1)_R$, $U(1)^3$, $U(1)^2U(1)_R$, 
$U(1)U(1)_R^2$, 
$U(1)_R^3$,
$U(1)$ and $U(1)_R$ has been noted 
in \cite{KSS}, and can be extended to the discrete symmetries 
$SU(F)^2Z_{3F}$, $Z_{3F}$, $Z_{3F}^3$, $U(1)^2Z_{3F}$, $U(1)Z_{3F}^2$, 
$U(1)_R^2Z_{3F}$,
$U(1)_RZ_{3F}^2$, $U(1)U(1)_RZ_{3F}$ as well. 
We will argue 
that the theory is indeed s-confining, that is it is described by these 
gauge singlets everywhere on the moduli space. For this in the 
next section we will 
 analyze the classical limit of the theory, that is the classical constraints 
satisfied by these operators. This will completely determine the
form of the confining superpotential.\footnote{It has been recently argued
\protect\cite{syzygies}, that a sufficient and necessary condition for the
continuous 't Hooft anomaly matching conditions to be satisfied by the
holomorphic gauge invariants is that the classical constraints be derivable
from a superpotential. Below we find this superpotential whose existence
is guaranteed by the above quoted theorem, and argue that this is indeed
the full confining superpotential of the theory.}  
Then we will show that 
instanton effects in the dual magnetic theory indeed do generate this
confining superpotential term.

\subsubsection{Analysis in the Electric Theory}  

In this section we will give a method to analyze the classical 
constraints of this theory and show the superpotential which 
can reproduce these constraints.
First we note that the analysis is different than in the 
ordinary s-confining theories, since there is a tree-level superpotential 
present in this theory, and the resulting $F$-flatness conditions have to be 
taken into account when analyzing the constraints.
The tree-level superpotential is ${\rm Tr}X^3$, and the resulting 
$F$-flatness condition is 
\begin{equation} X^2-\frac{1}{N} {\rm Tr} X^2 =0,\end{equation}
that is $X^2 \propto 1$. One can use the complexified gauge group to 
transform $X$ to a Jordan normal form, {\it e.g.}\/,
\begin{equation} X=\left( \begin{array}{cccccccc}
a & 1 \\
& a & 1 \\
& & a \\
& & & b & 1 \\
& & & & b \\
& & & & & c \\
& & & & & & \ddots \\
& & & & & & & z \end{array} \right)
\end{equation}
In the case of a completely diagonal $X$, 
one can easily see that $a=b=\cdots =z=0$ as follows.  The $F$-flatness
condition $X^2\propto 1 $ forces all the diagonal elements to be $\pm v$ while
there are odd number ($2F-1$) of diagonal elements, and hence their sum 
can not be zero, unless 
$v=0$. The diagonal elements $a,b, \cdots ,z$ have to vanish
in the case of the general Jordan normal form as well, since the diagonal 
elements 
will still be $a^2,a^2,a^2,b^2,b^2,c^{2},\cdots , z^2$.  Thus the only 
possible 
form of the matrix $X$ is diagonal entries with the eigenvalue 0 and 
a certain number of non-vanishing $2\times 2$ 
blocks of the form
\begin{equation} \left( \begin{array}{cc} 0 & 1\\ & 0 \end{array} \right) 
\end{equation}
to guarantee $X^{2} = 0$.  The most general form then is:
\begin{equation}
X=\left( \begin{array}{cccccccc}
0 &  \\
& \ddots &  \\
& & 0 \\
& & & 0 & 1 \\
& & & & 0 \\
& & & & & \ddots  \\
& & & & & & 0 & 1 \\
& & & & & & & 0 \end{array} \right) \label{eq:Xgeneral}
\end{equation}
This classical analysis immediately justifies the fact that ${\rm Tr} X^2$
(and all other invariants of the form  ${\rm Tr} X^p$)
is not among the confining degrees of freedom, since for all configurations
satisfying $F$-flatness ${\rm Tr} X^2=0$. 

Next we will identify the classical constraints among the 
gauge-invariant polynomials $M_1$, $M_2$, $B$ and $\bar{B}$.  For 
this, we introduce $F$ {\it dressed flavors}\/, $XQ,X\bar{Q}$, in 
addition to the original $F$ flavors $Q,\bar{Q}$.  Thus we consider 
the enlarged flavor space ${\cal Q}=(Q,XQ)$, $\bar{\cal Q}=(\bar{Q}, 
X\bar{Q})$.  Treating all $2F$ ``flavors'' independently, we find the 
same classical constraints as in an $SU(2F-1)$ theory with the $2F$ 
flavors.  The classical constraints among meson and baryon operators 
in this case are well-known from the analysis of the $SU(N)$ theories 
with $N+1$ flavors \cite{Seiberg}.\footnote{This theory is 
s-confining.  Note, however, that this analysis is strictly classical 
and it may or may not be 
a coincidence that both of these theories are s-confining.}

The meson matrix of the theory with dressed flavors is given by
\begin{equation} {\cal M} = \bar{\cal Q} {\cal Q} =
\left( \begin{array}{cc} \bar{Q}Q & \bar{Q} XQ \\
\bar{Q} XQ & \bar{Q}X^2 Q\end{array} \right). \end{equation}
However, we know that due to the $F$-flatness conditions $X^2=0$, and
we obtain
\begin{equation} 
\label{dressed}
{\cal M}=\left( \begin{array}{cc}
M_1 & M_2 \\
M_2 & 0 \end{array} \right).\end{equation}
Similarly, we can construct the baryons for the enlarged flavor space:
\begin{equation} {\cal B}=(Q^{F-1}(XQ)^F, Q^F(XQ)^{F-1}), \qquad
\bar{{\cal B}}= (\bar{Q}^{F-1}(X\bar{Q})^F, \bar{Q}^F(X\bar{Q})^{F-1}). \end{equation}
The second components of ${\cal B},\bar{{\cal B}}$ 
are $B,\bar{B}$ of (\ref{spectrum}),
and we will argue that the first components vanish due to the 
$F$-flatness conditions. This is because $X$ has at most $F-1$ non-vanishing
elements (otherwise $X^2$ would not be vanishing, since $X$ is a 
$2F-1$ by $2F-1$ matrix; see Eq.~(\ref{eq:Xgeneral})), and the color 
index contraction in $Q^{F-1}(XQ)^F$ yields a vanishing result due to 
the antisymmetry in color.
This also explains why the baryons formed this way are not part of the
confining spectrum.
Thus
\begin{equation} {\cal B}=(0,B), \qquad \bar{{\cal B}}=(0,\bar{B}). \end{equation}
We know that in the enlarged flavor space the classical constraints are
given by
\beq
{\cal M}_{ij}{\cal B}^j=0, \quad \bar{{\cal B}}^i{\cal M}_{ij}=0,\quad 
\bar{{\cal B}}^i{\cal B}^j={\rm cof} {\cal M}^{ij},
\eeq
where the cofactor of a $p$ by $p$ matrix $A$ is defined as
\begin{equation} ({\rm cof}\, A)^{ij}=\frac{1}{(p-1)!}
\epsilon^{i{i_2}{i_3}\cdots {i_p}}
\epsilon^{j{j_2}{j_3}\cdots {j_p}}A_{{i_2}{j_2}}A_{{i_3}{j_3}}\cdots 
A_{{i_p}{j_p}}=\frac{\partial {\rm det}\, A}{\partial A_{ij}}.\end{equation}
Written in terms of the confined variables $M_1$, $M_2$, $B$ and $\bar{B}$ 
these constraints read:
\begin{eqnarray}
&{M_2}_{ij}B^j=0,& \nonumber \\
&  \bar{B}^i {M_2}_{ij} =0,& \nonumber \\
&\left( \begin{array}{cc} 
0 & 0 \\
0 & \bar{B}^i B^j \end{array} \right) = \left( \begin{array}{cc} 
0 & {\rm det}M_2 {\rm cof} M_2^{ij} \\
 {\rm det}M_2 {\rm cof} M_2^{ij} & (M_1 {\rm cof} M_2)  {\rm cof} M_2^{ij}
\end{array} \right).&
\end{eqnarray}
The superpotential which reproduces these classical constraints is given 
by
\beq
\label{suppot}
W=\frac{1}{h^{2F-1}\Lambda^{6F-4}}
\left( \bar{B}M_2 B -{\rm det} M_2 (M_1 {\rm cof}M_2)\right) ,
\eeq
where $\Lambda$ is the dynamical scale of the original confining
$SU(2F-1)$ gauge group. Note that, unlike in the usual s-confining theories
\cite{Seiberg,s-conf}, one has the two-instanton factor appearing in the
confining superpotential, rather than the 1-instanton factor $\Lambda^{3F-2}$.

One should ask the question whether the fact that (\ref{suppot}) reproduces
the classical constraints in itself is enough evidence for it being the 
full confining superpotential. The answer is no for the following reason:
one wants to obtain the classical limit when the expectation values of the
fields are big, $\langle \Phi \rangle \gg \Lambda$. This means that the
highest powers in $1/\Lambda$ in the superpotential have to reproduce 
the classical constraints, thus (\ref{suppot}) can not contain terms 
of higher order in $1/\Lambda$. However, since (\ref{suppot}) has only a 
term containing the 2-instanton factor, it is in principle possible 
that an additional term proportional to an integer power of the 
one-instanton factor
$1/\Lambda^{3F-2}$ is present in the
superpotential. We show, however, that this is not possible, if all 
fields appear with positive powers, and thus (\ref{suppot}) is indeed
the full superpotential. As explained above, the only possible additional 
term should be proportional to $1/\Lambda^{3F-2}$. Then the form of the extra
piece in the superpotential in terms of the high energy fields is fixed by the
global symmetries to be
\begin{equation}
\label{extra}
\frac{1}{\Lambda^{3F-2}} (Q^F X^{1+F}\bar{Q}^F).
\end{equation}
However, it is impossible to write this combination of fields in terms of
the confining spectrum (\ref{spectrum}). The reason is that due to the $U(1)$ 
baryon number $B$ and $\bar{B}$ would have to appear with the same power. Thus
we would have to use a combination of $(B\bar{B})$, $M_1$ and $M_2$ to
obtain (\ref{extra}). This is however impossible, since $(B\bar{B})$, $M_1$ and
$M_2$ contain more or equal number of $(\bar{Q}Q)$'s than $X$'s, while
(\ref{extra}) contains more $X$'s than $(\bar{Q}Q)$'s. Thus we conclude that
(\ref{suppot}) is indeed the full confining superpotential.

This conclusion however might change if additional tree-level superpotential
terms (other than $M_1$) are added to the theory, for example a term
proportional to $M_2$. The reason is that in this case the 
classical constraints arising from the $F$-terms are modified, and the
analysis presented above has to be changed, which invalidates the form of the
dressed meson matrix ${\cal M}$ and the dressed baryons ${\cal B},
\bar{{\cal B}}$ of (\ref{dressed}). The most general confining superpotential
for this theory would incorporate the dependence on the 
coupling constants of all possible tree-level superpotential terms, and 
reduce to (\ref{suppot}) in the limit where all couplings other than ${\rm 
Tr}\, X^3$ are turned off.
We leave the determination of the most general superpotential for future
work, and note only that (\ref{suppot}) can be used when a superpotential
proportional to $M_1$ is added to the theory. This is because in this case
the $F$-term equation for $X$ is not affected by the presence of 
the additional tree-level superpotential term, and therefore the form 
of the dressed meson ${\cal M}$ remains unchanged. This perturbation is
what we will use later to integrate out a flavor from the theory.

\subsubsection{The superpotential from the Magnetic Theory} 

We will now argue, that the superpotential (\ref{suppot}) 
is indeed generated in the dual
magnetic theory, when integrating out a flavor. We start with a theory 
which has one more flavors ($F+1$) than the s-confining case, 
and $N$ is still given by $2F-1$. The theory thus has a magnetic dual 
in terms of an
$SU(3)$ gauge group. The dual theory is given by
\begin{equation} \begin{array}{c|c|cccc}
& SU(3) & SU(F+1) & SU(F+1) & U(1) & U(1)_R \\ \hline
Y & Adj & 1 & 1 & 0 & 2/3 \\
q & \Yfund & \overline{\Yfund} & 1 & 1 & 1-\frac{2}{F+1} \\
\bar{q} & \overline{\Yfund} & 1 & \overline{\Yfund} & -1 &  1-\frac{2}{F+1} \\
M_1 & 1 & \Yfund & \Yfund & 0 & 2-\frac{4(2F-1)}{3(F+1)}\\
M_2 & 1 & \Yfund & \Yfund & 0 & \frac{8}{3}-\frac{4(2F-1)}{3(F+1)}\\
\end{array}
\end{equation}
\begin{equation} W_{magn}= -h {\rm Tr} Y^3 + \frac{h}{\mu^2}\left( 
M_1 \bar{q}Yq +M_2 \bar{q}q \right) .\end{equation}
When integrating out a flavor in order to arrive at the s-confining case, we 
add a term $m Q_{F+1}\bar{Q}_{F+1}$ to the electric theory, which modifies the
magnetic superpotential to
\begin{equation} W_{magn}= -h {\rm Tr} Y^3 + \frac{h}{\mu^2}\left( 
M_1 \bar{q}Yq +M_2 \bar{q}q\right)  +m {M_1}_{F+1,F+1}.\end{equation}
The equation of motion with respect to ${M_1}_{F+1,F+1}$ forces an expectation
value to $\bar{q}_{F+1}Yq_{F+1}$, breaking the gauge group completely.
The expectation values which satisfy $D$- and $F$-flatness are given by:
\beq
\langle \bar{q} \rangle =(v,0,0), \; \langle Y \rangle =
\left( \begin{array}{ccc} 0&v\\
&0\\
&&0\end{array} \right), \; \; \langle q \rangle = \left( \begin{array}{c}
0\\v\\0 \end{array} \right).
\eeq
These VEV's, while breaking $SU(3)$ completely, give masses either 
through the superpotential or through the $D$-terms to all components of
$Y$, the elements of the last row and column of $M_1$ and $M_2$,
$q_{F+1}$ and $\bar{q}_{F+1}$, and the first and second components of all other
$q$'s and $\bar{q}$'s. The remaining components of $q$ and $\bar{q}$ can
be identified with $B$ and $\bar{B}$ of the confining theory. The 
tree-level superpotential $M_2 \bar{q}q $
will then result in the term $\bar{B}M_2B$. 
The remaining question is how to obtain the term 
${\rm det}\; M_2 (M_1 {\rm cof}M_2)$. This will be generated by a 
2-instanton effect in the completely broken magnetic group. 
The 't Hooft vertex for the 2-instanton is given by:
\begin{equation} \lambda^{12} \tilde{q}^{2F+2}\tilde{\bar{q}}^{2F+2} \tilde{Y}^{12} 
\Lambda_{magn}^{10-2F},\end{equation}
where $\lambda$ is the gaugino, and the other fields denote the fermionic 
components of the given chiral superfields. 
We will use the tree-level superpotential couplings and the 
gaugino-fermion-scalar vertices to convert this to a term in the 
superpotential~\cite{CM}. First we use the 
$q^*\lambda \tilde{q}$ and $\bar{q}^* \lambda \tilde{\bar{q}}$ vertices
three times and the $Y^* \lambda \tilde{Y}$ vertex once to convert 
the 't Hooft vertex to
\begin{equation} \Lambda_{magn}^{10-2F} \tilde{Y}^{11} Y^* \lambda^5 \tilde{q}^{2F-1}
\tilde{\bar{q}}^{2F-1} q^{*\; 3} \bar{q}^{*\; 3}. \end{equation}
Next we use $\frac{h}{\mu^2} M_1\bar{q}Yq$ superpotential coupling to obtain
\begin{equation}  \Lambda_{magn}^{10-2F} \tilde{Y}^{11} \lambda^5 \tilde{q}^{2F-2}
\tilde{\bar{q}}^{2F-2} q^{*\; 3} \bar{q}^{*\; 3} M_1 \frac{h}{\mu^2}.\end{equation}
Then we use the $\frac{h}{\mu^2} M_2 \bar{q}q$ term twice to obtain
\begin{equation} \Lambda_{magn}^{10-2F} \tilde{Y}^{11} \lambda^5 \tilde{q}^{2F-3}
\tilde{\bar{q}}^{2F-3} q^{*\; 2} \bar{q}^{*\; 2} M_1 \tilde{M}_2^2
\left(\frac{h}{\mu^2}\right)^{3}.\end{equation}
Finally we use the $\frac{h}{\mu^2} 
M_2 \bar{q}q$ superpotential term $2F-3$ times,
the $-h Y^3$ term 3 times and the $Y^* \lambda \tilde{Y}$ coupling five times
to obtain the term
\begin{equation}  \Lambda_{magn}^{10-2F}  M_1 \tilde{M}_2^2 M_2^{2F-3}
\langle Y \rangle ^3 \langle q^* \rangle^2 \langle \bar{q}^* \rangle^2
\langle Y^* \rangle^5 \frac{h^{2F+3}}{\mu^{4F}} . \end{equation}
Now we substitute the expectation values $v$ for $q, \bar{q}$ and $Y$ to 
obtain 
\begin{equation}  \Lambda_{magn}^{10-2F} M_1 \tilde{M}_2^2 M_2^{2F-3} v^3 
v^{*\; 9}\frac{h^{2F+3}}{\mu^{4F}}. \end{equation}
In order for the superpotential to be holomorphic, the additional factors of
$vv^*$ appearing from the integral over the instanton size have to cancel the
dependence on $v^*$\cite{CM}. Therefore one expects that the instanton 
integral results in an additional factor of $(vv^*)^{-9}$. Thus, we obtain
that the two-instanton in the completely broken $SU(3)$ group
gives a contribution to the superpotential of the form
\begin{equation}  \frac{h^{2F+3}\Lambda_{magn}^{10-2F}}{\mu^{4F}v^6}
{\rm det} \hat{M}_2 (\hat{M}_1 {\rm cof} \hat{M_2}), \end{equation}
where $\hat{M}_1$ and $\hat{M}_2$ are the meson operators for the theory 
with one less flavor. Let us check that the coefficient is indeed the
two-instanton factor of the electric theory as expected from (\ref{suppot}).
The matching of scales between the electric and magnetic theories is given by
\begin{equation} \Lambda_{el}^{3F-3} \Lambda_{magn}^{5-F}=\left(\frac{\mu}{h}\right)^{2F+2}.\end{equation}
The expectation 
value $v$ is given by $v^3=\mu^2 m/h$, thus we obtain that the superpotential 
term is (leaving the hats off)
\begin{equation} \frac{1}{h^{2F-1}m^2 \Lambda_{el}^{6F-6}} {\rm det} M_2 (M_1 {\rm cof}M_2).\end{equation}
Taking into account the scale matching in the electric theory
$\Lambda_{el}^{3F-3}m =\tilde{\Lambda}_{el}^{3F-2}$ we obtain exactly 
the second term in (\ref{suppot}) from this two-instanton effect.
Thus we conclude that the $2F-N=1$ theory is described by the
superpotential (\ref{suppot}), which correctly reproduces the 
classical constraints of the theory, and which can be shown to arise from
the dual magnetic theory when integrating out a flavor.

\subsubsection{Integrating out Flavors}

Using the results from the previous section we can obtain results for theories
with fewer number of flavors. Contrary to SUSY QCD and all other s-confining
theories which confine without the presence of a tree-level superpotential,
the theories with one less flavor does not yield a theory with a 
quantum modified constraint, instead it will result in a theory with a 
dynamically generated Affleck-Dine-Seiberg-type superpotential \cite{ADS}.
One can expect this by realizing, that the dual gauge group is $SU(2F-N)$,
thus integrating out a single flavor will result in breaking two colors 
instead of just one (or $k$ colors for a superpotential ${\rm Tr}\, X^{k+1}$).
Here we show how to integrate out a single flavor from the confining theory
presented in the previous section. 

Adding a mass term to one  flavor results in the superpotential
\begin{equation}
W=\frac{1}{h^{2F-1}\Lambda^{6F-4}} 
\left( \bar{B}M_2B-{\rm det}\, M_2 (M_1 {\rm cof}\,M_2)\right) +m(M_1)_{FF}.
\end{equation}
The $\bar{B}$, $B$ equations of motion just set the baryons to zero. The
$(M_1)_{FF}$ equation of motion gives 
\begin{equation}
(M_2)_{FF}=-\frac{mh^{2F-1}\Lambda^{6F-4}}{({\rm det}\, \tilde{M}_2)^2},
\end{equation}
where $\tilde{M}_2$ is the $M_2$ meson matrix for the theory with one less
flavors. The $\bar{B}M_2B$ piece and the pieces which contain
$(M_1)_{FF}$ of the superpotential are set to zero, so the only remaining piece
can be written as
\begin{equation}
W_{eff}=\frac{1}{h^{2F-1}\Lambda^{6F-4}}(M_2)_{FF}
({\rm det}\, \tilde{M}_2)\tilde{M}_1
({\rm cof}\, \tilde{M}_2)(M_2)_{FF}
=m^2 h^{2F-1}\Lambda^{6F-4} 
\frac{\tilde{M}_1{\rm cof}\, \tilde{M}_2}{({\rm det}\, \tilde{M}_2)^3}.
\end{equation}
Using the scale matching relation $m\Lambda^{3F-2}=\tilde{\Lambda}^{3F-1}$,
we obtain that the dynamically generated superpotential is given by
\begin{equation} W_{ADS}=\frac{h^{2F-1}\Lambda^{6F-2} (M_1 {\rm cof}\, M_2)}{ ({\rm det}\, M_2)^3}.
\end{equation}
This has the right quantum numbers to be a two-instanton effect in the 
$SU(2F-1)$ theory with $F-1$ flavors and an adjoint. Note, that this 
superpotential (contrary to the confining superpotentials) does vanish 
in the $h\to 0$ limit. Similarly, one can
integrate out further flavors to obtain the dynamically generated 
superpotentials for the theories with fewer flavors, which we leave 
as an exercise to the reader.

\subsection{The $SU$ theories for $k>2$}

Next we discuss the theories with $k>2$, with the 
superpotential $W=h{\rm Tr}X^{k+1}$.\footnote{Now the coupling $h$ is 
dimensionful.}  First of all, one can 
obtain the confining spectrum similar to (\ref{spectrum}) which satisfies the
't Hooft anomaly matching conditions for arbitrary $k$. This 
spectrum is given in the table below for $N=kF-1$.

\begin{eqnarray}
\label{spectrum2}
\begin{array}{c|c|ccccc}
& SU(N) & SU(F) & SU(F) & U(1) & U(1)_R & Z_{(k+1)F} \\ \hline
X & Adj & 1 & 1 & 0 & \frac{2}{k+1} & F\\
Q & \Yfund & \Yfund & 1 & 1 &  1-\frac{2(kF-1)}{(k+1)F}& -(kF-1)\\
\bar{Q} & \overline{\Yfund} & 1 & \Yfund & -1 &   1-\frac{2(kF-1)}{(k+1)F}&
-(kF-1) \\ \hline \hline 
M_i &  & \Yfund & \Yfund & 0 & \frac{2(2+F(1+i-k))}{F(1+k)}
& 2+F(i+2)\\
B &  & \overline{\Yfund} & 1 & kF-1 &  
\frac{F+kF-2}{F+Fk}& -1-\frac{k(k+1)F^2}{2}\\
\bar{B} &  & 1 & 
\overline{\Yfund} & -kF+1 
&  \frac{F+kF-2}{F+Fk} & -1-\frac{k(k+1)F^2}{2}\\
\end{array}, \nonumber \\ 
\nonumber \\ \hspace*{-30pt} \end{eqnarray}
where $i=1,\cdots ,k$ and the generalized mesons and baryons are defined by
\begin{eqnarray}
&&M_i=(\bar{Q}X^{i-1}Q) ,\nonumber \\
&&B=(Q^F(XQ)^F\cdots (X^{k-1}Q)^{F-1}) ,\nonumber \\
&&\bar{B} =(\bar{Q}^F (X\bar{Q})^F\cdots (X^{k-1}\bar{Q})^{F-1}).
\end{eqnarray}
This spectrum satisfies all the 't Hooft anomaly matching conditions 
for $SU(F)^3$, 
$SU(F)^2U(1)$, $SU(F)^2U(1)_R$, $U(1)^3$, $U(1)^2U(1)_R$, 
$U(1)U(1)_R^2$, $U(1)_R^3$, $U(1)$, $U(1)_R$,
$SU(F)^2Z_{(k+1)F}$, $Z_{(k+1)F}$, $Z_{(k+1)F}^3$, $U(1)^2Z_{(k+1)F}$, 
$U(1)Z_{(k+1)F}^2$, $U(1)_R^2Z_{(k+1)F}$,
$U(1)_RZ_{(k+1)F}^2$, $U(1)U(1)_RZ_{(k+1)F}$.
For $k=1$, the spectrum correctly 
reproduces the s-confining spectrum of the
$SU(F-1)$ theory with $F$ flavors.
However, in order to establish that the theory is in the 
confining phase, one has to show that these are the only flat 
directions of the theory.  We will show that this is indeed the case 
in a rather non-trivial manner;
all invariants of the form ${\rm Tr}\, X^p$ are lifted either by the
$F$ and $D$-flatness conditions or by non-perturbative quantum effects.

The $F$-flatness condition for the tree-level superpotential
${\rm Tr}\, X^{k+1}$ is
\beq
X^k-\frac{1}{N} {\rm Tr}\, X^k=0.
\eeq
Thus $X^k \propto 1$, and the diagonal elements of $X$ in the Jordan normal
form must be $k$-th roots of unity.  On the other hand, the
sum of the diagonal elements has to vanish since ${\rm Tr}\, X=0$.  
If $k$ is prime, we 
cannot have $kF-1$ eigenvalues (all $k$-th root of unity) 
summing up to zero, and all the 
eigenvalues must vanish.  This proves that the 
operators ${\rm Tr}\, X^{p}$  are all lifted.

However, one can find classical flat directions where only the field $X$ 
has an expectation value when 
$k$ is not prime.  This would lead to additional branches of the theory 
and to invariants of the form ${\rm Tr}\, X^{p}$ in addition
to those listed above. For example, in the case $N=5$, $k=6$, $F=1$,
there is a direction 
\begin{equation} X=v\left( \begin{array}{cccccc}
1\\&1\\&&-1\\&&&\omega\\&&&&\omega^2 \end{array} \right), \end{equation}
where $\omega=\frac{-1+i\sqrt{3}}{2}$. This direction satisfies
the $F$-term condition $X^6-\frac{1}{5} {\rm Tr}\, X^6=0$ since all elements
are sixth roots of unity, and also ${\rm Tr}\, X=0$. This would mean
that in addition to the operators listed above one would need to include
${\rm Tr}\, X^2$, ${\rm Tr}\, X^3$, ${\rm Tr}\, X^4$, and ${\rm Tr}\, X^6$ 
into the spectrum, which would give rise to a
Coulomb branch and the theory would likely not be confining at the
origin. This is however not the case.
Under the unbroken $SU(2)$ gauge group left by the above flat 
direction, we have one flavor of quarks, which leads to the
Affleck--Dine--Seiberg superpotential.  Therefore, this classical flat 
direction is lifted quantum mechanically and is removed from the 
quantum moduli space.

The same mechanism lifts all classical flat directions where only 
$X$ has an expectation value, as 
can be proven below.  If we have $p_1$ diagonal entries in $X$ which 
are the first $k$-th root of unity, $p_2$ of the second one etc., the 
gauge group is broken to $SU(p_1)\times SU(p_2)\times \cdots \times 
SU(p_k)\times U(1)^{k-1}$, with each $SU(p_i)$ factor having $F$ 
flavors.  We will show that such directions are lifted by quantum 
effects.  This happens if there is an ADS-type superpotential in any 
of the $SU(p_i)$ factors, that is if $p_i>F$ for some $i$.  Assume the 
contrary, that is $p_i\leq F$ for all $i$.  This is only possible if 
for example $p_1=p_2=\cdots =p_{k-1}=F, p_k=F-1$ or its permutations, 
since the size of the gauge group is $kF-1$.  However, in this case 
the adjoint is not traceless, which is a contradiction, and hence 
$p_i>F$ for at least one $i$.  Therefore configurations of the 
$X$ which are $F$-flat and not lifted by quantum effects have only 
vanishing diagonal elements.  All dangerous classical flat directions 
leading to Coulomb branches are lifted by quantum effects 
(all operators ${\rm Tr} X^p=0$), and the theory is indeed confining 
for any value of $k$.  The most general $X$ configuration, which can 
be made $D$-flat together with $Q$ and $\bar{Q}$, is then given in 
Jordan normal form with blocks of the form
\begin{equation}
	\left( \begin{array}{cccccc}
		0 & 1 & & & &\\
		& 0 & 1 & & &\\
		& & & \ddots & &\\
		& & & & 0 & 1\\
		& & & & & 0
		\end{array} \right)
\end{equation}
where each of the blocks is at most $k\times k$ to satisfy the 
$F$-flatness condition $X^{k}=0$.

The confining superpotential
can be fixed similarly to the case of $k=2$. One again considers the 
dressed flavors
${\cal Q}=(Q,XQ,X^2Q,\cdots ,X^{k-1}Q)$ and 
$\bar{\cal Q}=(\bar{Q},X\bar{Q},X^2\bar{Q},\cdots,X^{k-1}\bar{Q})$. 
Then we can construct the mesons 
${\cal M}=\bar{\cal Q}{\cal Q}$ and baryons ${\cal B}$
for these dressed flavors:
\begin{eqnarray}
{\cal M}&=&\left( \begin{array}{ccccc}
M_1 & M_2 & M_3 & \cdots & M_k \\
M_2 & M_3 & \cdots & M_k & 0\\ 
M_3 & \cdots & M_k & 0 & 0\\
\vdots & & & & \vdots \\
M_k & 0 & \cdots & & 0 \end{array} \right), \nonumber \\
{\cal B}&=&(0,0,\cdots ,B), \nonumber \\
\bar{\cal B}&=&(0,0,\cdots ,\bar{B}).
\end{eqnarray}
The fact that other components in ${\cal B}$, $\bar{\cal B}$ vanish 
can be shown based on the same argument for the $k=2$ case in the 
previous section.  
We then rewrite  the classical constraints ${\cal B}\bar{\cal B}={\rm 
cof}\, {\cal M}$ and 
${\cal BM}=\bar{\cal B}{\cal M}=0$ in terms of $M_i$, $B$ and $\bar{B}$. 
The latter two conditions
are satisfied if the term $BM_k\bar{B}$ is present in the superpotential,
while the ${\cal B}\bar{\cal B}={\rm cof}\, {\cal M}$ implies that the full superpotential 
is of the form
\begin{eqnarray}
W&=& \frac{1}{h^{kF-1}\Lambda^{k((2k-1)F-2)}}\left[
BM_k\bar{B}+ ({\rm det}\, M_k)^{k-1} (M_1 {\rm cof}\, M_k)+ \right.\nonumber \\
&& ({\rm det}\, M_k)^{k-2} (M_2 {\rm cof}\, M_k)(M_{k-1} {\rm cof}\, M_k)+
 ({\rm det}\, M_k)^{k-2} (M_3 {\rm cof}\, M_k) (M_{k-2} {\rm cof}\, M_k)+ 
\cdots
\nonumber \\
&& + \cdots + \nonumber \\
&& \left. \cdots + ({\rm det}\, M_k) (M_{k-1} {\rm cof}\, 
M_k)^{k-1}\right],
\end{eqnarray}
where we have not fixed the relative coefficients in the above superpotential.
The structure of the terms in the above superpotential is such that
${\rm det}\, M_k$ has to appear at least once in every term 
(except the first term $BM_k\bar{B}$) at least once. After the power of
${\rm det}\, M_k$ in a given term is fixed one has to add all possible
terms containing the appropriate number of $X$ and $Q,\bar{Q}$ fields to 
find the most general superpotential. 

Note that the overall dependence on the instanton-factor 
$\Lambda^{(2k-1)F-2}$ is that of the $k$-instanton.
For example, in the case $k=3$, the
superpotential (again without fixing the relative coefficients) has the 
form
\beq
\frac{1}{h^{3F-1}\Lambda^{15F-6}}
\left[ BM_3\bar{B} + ({\rm det}\, M_3)^2 (M_1 {\rm cof}\, M_3)+
 ({\rm det}\, M_3) (M_2 {\rm cof}\, M_3)^2\right].
\eeq

\section{Other Models\label{sec:other}}
\setcounter{equation}{0}
\setcounter{footnote}{0}

In this section we present other examples which  confine in the presence of a 
suitable tree-level superpotential, similarly to the theory presented in
the previous section. These examples are based on the dualities
presented in Refs.~\cite{Intr,LS,ILS}. For the first example we present both
the confining spectrum and the confining superpotential, and show how it is
obtained by a three-instanton effect by integrating out a flavor in the 
dual theory. For the remaining examples we give only the confining spectrum
satisfying the 't Hooft anomaly matching conditions, leaving the 
determination of the superpotentials for future work.

\subsection{$Sp$ with an adjoint and fundamentals}

This confining theory is based on the duality of Ref.~\cite{LS}, where
it is shown that an $Sp(2N)$ theory with an adjoint $X$ and $2F$ fundamentals
$Q$, and a superpotential $W_{tree}=h {\rm Tr} X^{2(k+1)}$ is dual to
$Sp(2\tilde{N})$ with an adjoint $Y$, $2F$ fundamentals $q$, and gauge singlet
mesons $M_i$, $i=0,...,2k$, with $\tilde{N}=(2k+1)F-N-2$. The 
confining case is obtained when $\tilde{N}=0$, that is for $N=(2k+1)F-2$.
The field content, symmetries and the confining spectrum are given in the 
table below:
\begin{equation}
\label{Sptable}
\begin{array}{c|c|ccc}
& Sp((2k+1)2F-4) & SU(2F) & U(1)_R & Z_{2(k+1)F} \\
Q & \Yfund & \Yfund & \frac{1-Fk}{F(k+1)} &-((2k+1)F-1)\\
X & \Ysymm & 1 & \frac{1}{k+1} & F \\ \hline \hline
M_{2i}=QX^{2i}Q &  & \Yasymm & \frac{2F+4Fi+4k-2Fk}{F+Fk}&
2(1+F(i+1))\\
M_{2j+1}=QX^{2j+1}Q &  & \Ysymm & \frac{4F+4Fj+4k-2Fk}{F+Fk}&2+3F+2Fj
\end{array}, \end{equation}
where $i=0,...,k$ and $j=0,...,k-1$.  It is straightforward to check 
that this particle content saturates all anomaly matching conditions,
including the discrete ones ($SU(2F)^3$, $SU(2F)^2U(1)_R$, $U(1)_R^3$, 
$U(1)_R$, $SU(2F)^2Z_{2(k+1)F}$, $U(1)_R^2 Z_{2(k+1)F}$, 
$U(1)_R Z_{2(k+1)F}^2$,
$Z_{2(k+1)F}$, $Z_{2(k+1)F}^3$). In addition, for $k=0$ it reproduces the
s-confining $Sp(2F-4)$ theory with $2F$ fundamentals.  

Next we determine the confining 
superpotential for $k=1$.  We need to first determine the classical
constraints in this theory, taking into account the $F$-flatness conditions.
Just like in the previous section, we consider an $Sp(6F-4)$ with
$6F$ dressed flavors ${\cal Q}=(Q, XQ, X^{2}Q)$.\footnote{The indices 
are contracted as $X^{ij}J_{jk}Q^{k}$ etc with the symplectic matrix 
$J$.} 
The classical constraints were given in Ref.~\cite{IntPoul}:
$\epsilon_{a_1a_2\cdots a_{6F}} {\cal M}^{a_3a_4} \cdots 
{\cal M}^{a_{6F-1}a_{6F}}=0$,
$a_1,a_2=1,\cdots ,6F$. In our case, the meson matrix ${\cal M}$ is given by
\begin{equation} {\cal M}=\left( \begin{array}{rcc}
M_0 & M_1 & M_2 \\
-M_1&M_2&0\\
M_2&0&0\\ \end{array} \right),\end{equation}
which already incorporates the $F$-flatness condition. In the following, we
will only consider the case of 
$F=1$, that is $SU(2)$ with an adjoint and a single flavor, and a 
superpotential $h {\rm Tr} X^4$. In this case one can easily see that 
the tree-level superpotential indeed lifts the Coulomb branch of the
theory. This is because the gauge group is just $SU(2)$, and the $h {\rm Tr} 
X^4$ superpotential is nothing but $({\rm Tr} X^2)^2$. Then the equation of
motion yields $({\rm Tr} X^2) X=0$, from which ${\rm Tr} X^2=0$, thus the
mesons of (\ref{Sptable}) are indeed sufficient to describe the moduli space 
of the $F=1,k=1$ theory.  We believe that the ${\rm Tr}X^{2p}$ operators 
are lifted by the $F$-flatness conditions for any $k$ and $F$, even 
though we have not proven it.

The classical constraints for the $k=1,F=1$ theory are given by
\begin{equation}
{\rm Pf}\, M_2 =0, \; \; {\rm cof}\, M_1 {\rm Pf}\, M_2=0, \; \;
{\rm det}\, M_1-{\rm Pf}\, M_2 {\rm Pf}\, M_0=0,
\end{equation}
where of an antisymmetric $2N$ by $2N$ matrix $A$ is defined by
\begin{equation}
{\rm Pf}\, A=\frac{1}{N!2^N} \epsilon^{i_1i_2 \cdots i_{2N-1}i_{2N}}
A_{i_1i_2} \cdots A_{i_{2N-1}i_{2N}} =\sqrt{{\rm det}\, A}.
\end{equation}
Note that for the case considered ($F=1$)  ${\rm Pf}\, M_0={(M_0)}_{12}$,
$ {\rm Pf}\, M_2={(M_2)}_{12}$.
These constraints can be derived from the superpotential 
\begin{equation}
{\rm Pf}\, M_2 \left( {\rm det}\, M_1 -\frac{1}{2}{\rm Pf}\, M_2 
 {\rm Pf}\, M_0 \right).
\end{equation}
Note, however, that simply by dimensional reasons one can not have 
the instanton factor $\Lambda^3$ to be the only constant
appearing in the superpotential, and the
extra mass scale $h^{-1}$ of the tree-level superpotential must appear in the
confining superpotential as well. The correct form of the superpotential
for $F=1$ is given by
\begin{equation}
\label{confsup}
\frac{1}{h^{2}\Lambda^9}{\rm Pf}\, M_2 \left( {\rm det}\, M_1 -
\frac{1}{2}{\rm Pf}\, M_2 
 {\rm Pf}\, M_0 \right).
\end{equation}
One can check that this superpotential is invariant under all the global
symmetries of the theory, including the anomalous ones if appropriate 
charges are assigned to $h$ and to $\Lambda^3$. Note that this 
superpotential is indeed a three-instanton effect, and that it diverges if the
tree-level superpotential is turned off, signaling that the description is
valid only if a tree-level superpotential is present. Next we explain how
this superpotential is generated by a three-instanton effect
of the dual theory.
Consider the theory with one more flavors, that is $SU(2)$ with an adjoint $X$,
four fundamentals $Q$, and a superpotential $h{\rm Tr}\, X^4$.
The duality is given in the table below.
\begin{equation}
\begin{array}{c|c|c}
& SU(2) & SU(4) \\ \hline 
Q & \Yfund & \Yfund \\
X & \Ysymm & 1 \\ \hline \hline
& Sp(6) & SU(4) \\ \hline
q & \Yfund & \Yfund \\
Y & \Ysymm & 1 \\
M_0 & 1 & \Yasymm \\
M_1 & 1 & \Ysymm \\
M_2 & 1 & \Yasymm \end{array}.\end{equation}
The magnetic superpotential is given by
\begin{equation}
\label{magn}
W_{magn}=-h {\rm Tr}\, Y^4 + \frac{h}{\mu^2} \left(
M_0 qY^2q+M_1qYq+M_2q^2 \right).
\end{equation}
The matching of scales is given by 
\begin{equation}
\label{matching}
\Lambda_{el}^2\Lambda_{magn}^6=
\left( \frac{\mu}{h}\right)^4.
\end{equation}
Integrating out a flavor from the electric theory 
corresponds to adding the linear term $mM_0$ to the magnetic superpotential
(\ref{magn}), which forces a non-vanishing expectation value for
$qY^2q$, completely breaking the magnetic $Sp(6)$ gauge group. In order to
obtain the superpotential (\ref{confsup}), we consider the
3-instanton 't Hooft vertex of the broken magnetic $Sp(6)$ theory:
\begin{equation}
\tilde{Y}^{24} \lambda^{24}  \tilde{q}^{12}
\frac{\mu^{12}}{h^{12}\Lambda_{el}^{6}},
\end{equation}
where we have already used the scale matching (\ref{matching}).
We can use the  $Y^* \lambda \tilde{Y}$ vertex twice to convert this to
\begin{equation}
\tilde{Y}^{22} \lambda^{22} Y^{*\, 2} \tilde{q}^{12} 
\frac{\mu^{12}}{h^{12}\Lambda_{el}^{6}}.
\end{equation}
Next we use  the $\frac{h}{\mu^{2}}qY^2qM_0$ superpotential term once to get
\begin{equation}
\tilde{Y}^{22} \lambda^{22} M_0 \tilde{q}^{10} 
\frac{\mu^{10}}{h^{11}\Lambda_{el}^{6}}.\end{equation}
Now we use the $q^* \lambda \tilde{q}$ vertex twice to get 
\begin{equation} \tilde{Y}^{22} \lambda^{20} M_0  \tilde{q}^{8}
q^{*\, 2} \frac{\mu^{10}}{h^{11}\Lambda_{el}^{6}}, \end{equation}
and then the $\frac{h}{\mu^{2}}M_2 q^2$ superpotential coupling twice to get
\begin{equation} \tilde{Y}^{22} \lambda^{20} M_0  \tilde{q}^{6} \tilde{M}_2^2
\frac{\mu^{6}}{h^{9}\Lambda_{el}^{6}}.\end{equation}
Next we use the $-h Y^4$ superpotential coupling four times to get
\begin{equation}
\langle Y \rangle^8
\tilde{Y}^{14} \lambda^{20} \tilde{q}^{6} M_0  \tilde{M}_2^2
\frac{\mu^{6}}{h^{5}\Lambda_{el}^{6}}.\end{equation}
Now we use the remaining fermionic components to convert them to VEV's:
\begin{equation}
\langle Y \rangle^8 \langle Y^*\rangle^{14}  
\langle q^*\rangle^{6} M_0 \tilde{M}_2^2
\frac{\mu^{6}}{h^{5}\Lambda_{el}^{6}}.\end{equation}
The expectation values for $q$ and $Y$ are given by $v=(\mu^2 
m/h)^{\frac{1}{4}}$. Because the superpotential has to be holomorphic in $v$, the
integral over the instanton size has to give an additional factor of
$(vv^*)^{20}$ in the denominator. Thus the superpotential term will 
be 
\begin{equation} \frac{\mu^{6} M_0 \tilde{M}_2^2}{v^{12} h^{5}\Lambda_{el}^{6}}.
\end{equation}
Substituting $v^{12}=\mu^6 m^3/h^{3}$ and using the scale matching of the 
electric theory $m^3 \Lambda_{el}^6=\tilde{\Lambda}_{el}^9$ we get the
instanton generated superpotential term
\begin{equation} \frac{1}{h^{2}\tilde{\Lambda}_{el}^9} M_0 M_2^2,\end{equation}
which is one of the terms of (\ref{confsup}). The other term can be presumably
generated from the same three-instanton vertex by closing up the instanton
legs with different vertices.

\subsection{$Sp$ with a traceless antisymmetric tensor and fundamentals}
\label{subsec:SpA}
We present only the confining spectrum which satisfies the 't Hooft anomaly
matching conditions $SU(2F)^3$, $SU(2F)^2U(1)_R$, $U(1)_R^3$, $U(1)_R$,
$SU(2F)^2Z_{(k+1)F}$, $ U(1)_R^2 Z_{(k+1)F}$, $U(1)_R Z_{(k+1)F}^2$,
$ Z_{(k+1)F}^3$ and $Z_{(k+1)F}$ for 
$Sp(2k(F-2))$ with a traceless antisymmetric tensor 
$X$ (${\rm Tr}JX=0$), $2F$ fundamental fields $Q$ and a tree-level superpotential
${\rm Tr} X^{k+1}$.
\begin{equation}
\begin{array}{c|c|ccc}
& Sp(2k(F-2)) & SU(2F) & U(1)_R & Z_{(k+1)F} \\ \hline
X & \Yasymm & 1 & \frac{2}{k+1} & F\\
Q & \Yfund & \Yfund & -\frac{Fk-2k-F}{F(1+k)} & 1-(-2+F) k \\ \hline \hline
M_i=QX^iQ & & \Yasymm & \frac{2i}{1+k}-2\frac{Fk-2k-F}{F(1+k)} &2+F i+4 k-
2 F k\\
T_j=X^j & & 1 & \frac{2j}{k+1} & Fj \end{array},
\end{equation}
where $i=0,\cdots ,k-1$ and $j=2,\cdots ,k$. For $k=1$ this theory reduces to
the s-confining $Sp(2F-4)$ theory with $2F$ fundamentals described in
\cite{IntPoul}.

It is interesting to note that some of the $T_{j}$ operators may be 
missing from the classical flat directions.  For instance, for $k=3$, 
$F=3$, one can easily show that $T_{2}$ vanishes due to the 
$F$-flatness conditions.  There is however no contradiction,
since for this particular 
case, $T_{2}$ has $U(1)_R$ charge one 
and $Z_{12}$ charge 6.  Therefore a mass term 
$m (T_{2})^{2}$ is allowed in the superpotential, and it is 
removed from the moduli space.

One consistency check for confinement in this theory is to consider the
classical flat direction of the form
\begin{equation}
	X = i\sigma_{2} \otimes \left( \begin{array}{cccc}
		1 & & &\\ & \omega & &\\ & & \ddots &\\ & & &\omega^{k-1}
		\end{array} \right),
\end{equation}
where $\omega = e^{2i\pi/k}$, and every diagonal element in $X$ is 
multiplied by the $F-2$ dimensional unit matrix.
  This direction indeed satisfies the 
$F$-flatness condition $(JX)^{k} \propto 1$, where $J = i\sigma_{2} 
\otimes 1$ is the symplectic matrix, and corresponds to the 
polynomial $T_{k}={\rm Tr}X^{2k}$.  The theory is indeed s-confining along 
this direction because it leaves an $(Sp(2(F-2)))^{k}$ gauge group 
unbroken with $F$ flavors for each $Sp(2(F-2))$ factor, and it confines 
\cite{IntPoul}.  If $k$ is non-prime, $k=lm$, however, one may worry 
about the following direction
\begin{equation}
	X = i\sigma_{2} \otimes \left( \begin{array}{cccc}
		1 & & &\\ & \omega^{l} & &\\ & & \ddots &\\ & & &\omega^{l(k-1)}
		\end{array} \right),
\end{equation}
where again every diagonal element is multiplied by the $F-2$ dimensional
unit matrix.
This direction is lifted quantum mechanically because it leaves an
$(Sp(2l(F-2)))^{m}$ gauge group unbroken, and each of the 
$Sp(2l(F-2))$ factor develops the Affleck--Dine--Seiberg 
superpotential.

\subsection{$SO$ with an adjoint and vectors}

The spectrum which satisfies the anomaly matching conditions
$SU(F)^3$, $SU(F)^2U(1)_R$, $U(1)_R^3$, $U(1)_R$, $SU(F)^2 Z_{2(k+1)F}$,
$U(1)_R^2Z_{2(k+1)F}$, $ U(1)_R Z_{2(k+1)F}^2$, $Z_{2(k+1)F}^3$,
$Z_{2(k+1)F}$, $SU(F)^2 Z_{2F}$,
$U(1)_R^2Z_{2F}$, $ U(1)_R Z_{2F}^2$, $Z_{2F}^3$,
$Z_{2F}$, $U(1)_R Z_{2F}Z_{2(k+1)F}$, $ Z_{2F}^2Z_{2(k+1)F}$ and
$ Z_{2F}Z_{2(k+1)F}^2$ is given by
\[ \begin{array}{c|c|cccc}
& SO(N) & SU(F) & U(1)_R & Z_{2(k+1)F} & Z_{2F} \\ \hline
X & \Yasymm & 1 & \frac{1}{k+1} & F & 0 \\
Q & \Yfund & \Yfund & -\frac{1+Fk}{F(1+k)}& -1-F-2 F k & 1 \\ \hline \hline
M_{2i}=QX^{2i}Q & & \Ysymm & \frac{2i}{1+k}-\frac{2(1+Fk)}{F(1+k)} & 
-2+2 F (-1+i-2 k) & 2 \\
M_{2j+1}=QX^{2j+1}Q & & \Yasymm & \frac{1+2j}{1+k}-\frac{2(1+Fk)}{F(1+k)} 
& -2+F (-1+2 j-4 k) & 2 \\
B=Q^{F-1}X^{\frac{N-F+1}{2}} & & \overline{\Yfund} & \frac{1+F+Fk}{F(1+k)}
& 1+2F(1+k)-F^2(1+k) & F-1 
\end{array},
\]
where $i=0,\cdots ,k$, $j=1,\cdots ,k-1$, $N=(2k+1)F+3$, and the 
tree-level superpotential of the $SO(N)$ theory is ${\rm Tr}\, 
X^{2(k+1)}$.

For $k=0$ this spectrum reproduces that of the $SO(F+3)$ theory 
with $F$ flavor, which has a confining branch (of which this is the 
generalization for $k>0$), and a branch with a dynamically generated 
superpotential.  We assume that this multiple branch structure 
persists for the case with $k>0$, therefore the above spectrum 
describes only one of the possible branches of the theory.  Naively, 
for $k=0$ this spectrum does not exactly agree with the 
spectrum of \cite{SO}, since the baryon operator is $B=Q^{F-1}X^2$ 
here, while the baryon in \cite{SO} is $W_{\alpha}^2 Q^{F-1}$.  
However, in the presence of the tree-level mass term in the 
superpotential $M_{X}{\rm Tr}\, X^2$, these two operators can be 
identified using the chiral anomaly equation~\cite{discrete}.  The 
argument is the following.  We start from the part $X^{ij}X^{kl}$ 
in the operator $B$, where none of the indices $i,j,k,l$ are the same 
because they are contracted with an epsilon tensor.  We contract two 
$X$ fields via a one-loop triangle diagram with two external gauge 
fields.  This is the same calculation as the contribution of the 
Pauli--Villars field in the Konishi anomaly \cite{Konishi}, except 
that the gauge indices are not contracted between two fields.  The 
result is proportional to $\frac{1}{32\pi^{2}M_{X}}$.  The two gauge 
vertices in the triangle diagram must transform the indices of $X^{ij}$ to 
those of $X^{kl}$, and hence require $SO(N)$ generators $M^{ik}$ and 
$M^{jl}$, or $M^{il}$ and $M^{jk}$.  Therefore the resulting gauge 
fields have indices $W_{\alpha}^{ik}W^{\alpha jl} - W_{\alpha}^{il} 
W^{\alpha jk}$.  Now recall that all these indices were contracted 
with the epsilon tensor, and the above two terms give identical 
contributions as $-W_{\alpha}^{ij}W^{\alpha kl}$.  Therefore the net 
result is to replace $X^{ij}X^{kl}$ by $-W_{\alpha}^{ij}W^{\alpha kl}$ 
apart from numerical factors and $1/M_{X}$.

Note that 
in the spectrum  the operator $B$ could have been substituted by the
operator $W_{\alpha}^{2(k+1)} Q^{(k+1)F-1} X^{k(F(1+k)-2k-3)}$ 
without the modification of any of the continuous global anomalies, and even
the $k=0$ limit of this operator is correct. The only way to distinguish 
between this operator and the operator $B$ which is the correct 
confined degree of freedom is by considering the discrete anomaly matching
conditions~\cite{discrete}, 
which is only satisfied if one uses the operator $B$.

\subsection{$SO$ with a traceless symmetric tensor and vectors}
\label{subsec:SOS}
The spectrum which satisfies the anomaly matching conditions
$SU(F)^3$, $SU(F)^2U(1)_R$, $U(1)_R^3$, $U(1)_R$, $SU(F)^2 Z_{(k+1)F}$,
$U(1)_R^2Z_{(k+1)F}$, $ U(1)_R Z_{(k+1)F}^2$, $Z_{(k+1)F}^3$,
$Z_{(k+1)F}$, $SU(F)^2 Z_{2F}$,
$U(1)_R^2Z_{2F}$, $ U(1)_R Z_{2F}^2$, $Z_{2F}^3$,
$Z_{2F}$, $U(1)_R Z_{2F}Z_{(k+1)F}$, $ Z_{2F}^2Z_{(k+1)F}$ and
$ Z_{2F}Z_{(k+1)F}^2$ is given by
\begin{equation} \begin{array}{c|c|cccc}
& SO(N) & SU(F) & U(1)_R & Z_{(k+1)F} & Z_{2F}\\ \hline
X & \Ysymm & 1 & \frac{2}{k+1} & F & 0 \\
Q & \Yfund & \Yfund & -\frac{(-2-F+4k+Fk)}{F(1+k)} & -(N-2) & 1 
\\ \hline \hline 
M_j=QX^jQ & & \Ysymm & \frac{2jF-2(-2-F+4k+Fk)}{F(1+k)}& 
jF-2(N-2)& 2 \\
B
& & \overline{\Yfund} & \frac{-2+F+4k+Fk}{F(1+k)}
& \frac{2 + 8\,k - {F^2}k( 1 + k)  + 
  F( 2 - 4k - 6{k^2} )}{2} & (kF-1) \end{array},
\end{equation}
where $j=0,\cdots ,k-1$, $N=k(F+4)-1$ and the tree-level superpotential
of the $SO(N)$ theory is ${\rm Tr}\, X^{k+1}$, and the 
field content of $B$ is $ W_{\alpha}^{2k} Q^{kF-1} X^{\frac{Fk(k-1)}{2}+(k+1)^2} $. The contraction of the 
gauge indices in $B$ is presumably
\begin{equation} B=Q^F (XQ)^F (X^2Q)^F \cdots (X^{k-2}Q)^F (X^{k-1}Q)^{F-1} (W_{\alpha})^2
(XW_{\alpha})^2 \cdots (X^{k-1}W_{\alpha})^2.\end{equation}
For $k=1$ this theory reproduces again the $SO(F+3)$ theory with $F$ 
vectors~\cite{SO}, 
and therefore similar multi-branch structure is expected in this
case as well.

One may worry about a possible Coulomb branch along the direction
\begin{equation}
	X = v \, {\rm diag}\, (x_{1}, x_{2}, \cdots, x_{k(F+4)-1}).
\end{equation}
To satisfy the $F$-flatness condition $X^{k} \propto 1$, all the 
eigenvalues $x_{i}$ must be $k$-th root of unity.  The tracelessness 
of $X$ also imposes the condition $\sum_{i} x_{i} = 0$.  If $k$ is 
prime, one cannot satisfy the tracelessness condition with $k(F+4)-1$ 
eigenvalues which are all $k$-th roots of unity.  Therefore, there is 
no classical flat direction of this form.  If $k$ is not prime, one 
may find divisors of $k$ which sum up to $k(F+4)-1$; {\it i.e.}\/, 
$p_{1}, \cdots, p_{m}$ are all divisors of $k$ and $\sum_{j} p_{j} = 
k(F+4)-1$.\footnote{For instance, for $k=6$ and $F=1$, $k(F+4)-1=29$, $X$ can 
be given by repeating $(1,\omega, \omega^{2}, 
\omega^{3},\omega^{4},\omega^{5})$ four times, and the 
remaining five eigenvalues by $(1, \omega^{3}, 1, \omega^{2}, 
\omega^{4})$.}  Then using the quotients $q_{m} = k/p_{m}$, one can 
satisfy both the tracelessness and $F$-flatness with
\begin{equation}
	X = v {\rm diag}(1, \omega^{q_{1}}, \cdots, \omega^{(p_{1}-1)q_{1}},
	1, \omega^{q_{2}}, \cdots, \omega^{(p_{2}-1)q_{2}}, \cdots,
	1, \omega^{q_{m}}, \cdots, \omega^{(p_{m}-1)q_{m}}) ,
\end{equation}
where $\omega = e^{2i\pi/k}$.  This direction corresponds to the 
polynomials ${\rm Tr}X^{p_{j}}$.  It is, however, lifted quantum 
mechanically.  The proof is by contradiction.  Suppose the direction 
is not lifted by quantum effects.  Since there are $F$ vectors, the 
unbroken group should not contain a factor larger than $SO(F+4)$ to 
avoid Affleck--Dine--Seiberg superpotential.  On the other hand, there 
are $k(F+4)-1$ eigenvalues with only $k$ possibilities $1, \omega, 
\cdots, \omega^{k-1}$, and hence the only allowed case is repeating 
all of the above $k$-th roots of unity $F+4$ times except one of them 
repeated $F+3$ times.  Then, however, $X$ is not traceless, and 
hence the assumption is not correct.

\subsection{$SU$ with an antisymmetric flavor and fundamental flavors}

The spectrum which satisfies the anomaly matching conditions
$SU(F)^3$, $SU(F)^2 U(1)_X$, $SU(F)^2U(1)_B$, $SU(F)^2U(1)_R$,
$U(1)_X^3$, $U(1)_X$, $U(1)_B^3$, $U(1)_B$,  $U(1)_R^3$, $U(1)_R$,
$U(1)_X^2 U(1)_B$, $U(1)_X U(1)_B^2$, $U(1)_X^2 U(1)_R$, $U(1)_X U(1)_R^2$,
$U(1)_B^2U(1)_R$, $U(1)_B U(1)_R^2$, $U(1)_X U(1)_B U(1)_R$, \break
$SU(F)^2 Z_{2(k+1)F}$, $Z_{2(k+1)F}$, $Z_{2(k+1)F}^3$,
$U(1)_X^2 Z_{2(k+1)F}$, $U(1)_X Z_{2(k+1)F}^2$,   
$U(1)_B^2 Z_{2(k+1)F}$, \break $U(1)_B Z_{2(k+1)F}^2$,
$U(1)_R^2 Z_{2(k+1)F}$, $U(1)_R Z_{2(k+1)F}^2$,
$U(1)_X U(1)_B Z_{2(k+1)F}$, $U(1)_X U(1)_R Z_{2(k+1)F}$ and
$U(1)_BU(1)_R Z_{2(k+1)F}$ is given by
\begin{equation} 
\begin{array}{c|c|cccccc}
& SU(N) & SU(F) & SU(F) & U(1)_X & U(1)_B & U(1)_R & Z_{2(k+1)F} \\ \hline
X & \Yasymm & 1 & 1 & 1 & 0 & \frac{1}{k+1} & F \\
\bar{X} & \overline{\Yasymm} & 1 & 1 & -1 & 0 & \frac{1}{k+1} & F \\
Q & \Yfund & \Yfund & 1 & 0 & 1 & 1-\frac{N+2k}{F(k+1)} & 3+4 k+F \\
\bar{Q} & \overline{\Yfund} & 1 & \Yfund & 0 & -1 &  1-\frac{N+2k}{F(k+1)} 
 & 3+4 k+F \\ \hline \hline
M_j & & \Yfund & \Yfund & 0 & 0 & \frac{2+2Fj+4k-2Fk}{
F(k+1)} &  {\scriptstyle 2jF-2(N-2)}\\
P_r & & \Yasymm & 1 & -1 & 2 & 
\frac{2+F+4k-2Fk+2Fr}{F(1+k)} &  {\scriptstyle (2r+1)F-2(N-2)} \\
\bar{P}_r & & 1 & \Yasymm & 1 & -2 & 
\frac{2+F+4k-2Fk+2Fr}{F(1+k)} &   {\scriptstyle (2r+1)F-2(N-2)} \\
B & & \overline{\Yfund} & 1 & \frac{N-F+1}{2} & 
F-1 & \frac{-1+F-2k+Fk}{F(1+k)}&   {\scriptstyle -3-4k-F^2(1+k)} \\
\bar{B} & & 1 & 
\overline{\Yfund} & -\frac{N-F+1}{2} & 
-F+1 & \frac{-1+F-2k+Fk}{F(1+k)} & {\scriptstyle  -3-4k-F^2(1+k)} \\
T_i & & 1 & 1 & 0 & 0 & \frac{2i}{k+1} & {2 F i} \\
\end{array},
\end{equation}
where $j=0,\cdots ,k$, $r=0,\cdots ,k-1$, $i=1,\cdots ,k$, $N=(2k+1)F-4k-1$,
and the tree-level superpotential of the $SU(N)$ theory is 
${\rm Tr} (X\bar{X})^{k+1}$, and the composite operators are given by
\begin{eqnarray}
&& M_j=Q(X\bar{X})^j\bar{Q}, \nonumber \\
&& P_r= Q(X\bar{X})^r\bar{X}Q, \nonumber \\
&& \bar{P}_r=\bar{Q}(X\bar{X})^rX\bar{Q}, \nonumber \\
&& B=X^{\frac{N-F+1}{2}} Q^{F-1},\nonumber \\
&& \bar{B}=\bar{X}^{\frac{N-F+1}{2}} \bar{Q}^{F-1}, \nonumber \\
&& T_i=(X\bar{X})^i. 
\end{eqnarray}
For $k=0$ this theory reproduces the 
s-confining $SU(F-1)$ theory with $F$ flavors. Note that there
are no $P$ or $T$ operators for $k=0$.  

When $N=(2k+1)F-4k-1$ is even ({\it i.e.}\/, $F$ odd), there is an 
additional classical flat direction:
\begin{equation}
	X = i\sigma_{2} \otimes {\rm diag}(x_{1}, \cdots, x_{N/2}) v ,
	\qquad \bar{X}=0.
\end{equation}
This direction obviously satisfies the $F$-flatness 
$\bar{X}(X\bar{X})^{k} = (X\bar{X})^{k} X = 0$ for $k>0$, and 
corresponds to the gauge-invariant polynomial ${\rm Pf}X$. 
$D$-flatness requires $X^{\dagger} X \propto 1$ and hence all the 
eigenvalues have the same absolute value.  Moreover, a general $SU(N)$ 
gauge transformation can make all the eigenvalues to be equal, and hence 
$X = i\sigma_{2} \otimes 1_{N/2} v$, which leaves an $Sp(N)$ subgroup 
unbroken.  The low-energy theory then is an $Sp(N)$ gauge theory with 
an anti-symmetric tensor $\bar{X}$ with superpotential ${\rm 
Tr}\bar{X}^{k+1}$.  The particle content is precisely the same as the 
model discussed in Section~\ref{subsec:SpA} except for the trace 
part of $\bar{X}$.  The gauge group $N=(2k+1)F-4k-1$ is larger than 
the confining case for which $N=2k(F-2)$ discussed in 
Section~\ref{subsec:SpA} by $\Delta N = F-1\geq 2$, because  
asymptotic freedom of the original $SU(N)$ theory requires $F \geq 3$.  
Then the theory is expected to develop an Affleck--Dine--Seiberg type 
superpotential and the direction is removed from the quantum moduli 
space for $F\geq 4$.\footnote{The case $F=3$ is not fully understood. 
The low-energy $Sp(N)$ dynamics is expected to give a  
quantum modified moduli space, while the trace part of $\bar{X}$ 
interacts with the traceless part of $\bar{X}$ via  the 
tree-level superpotential.  It is 
suggestive that the ${\rm Pf}X$ and ${\rm Pf}\bar{X}$ operators can be 
added to the confining spectrum without spoiling the anomaly matching 
conditions only when $F=3$.  It is likely that the theory is still 
confining together with these operators.}

\subsection{$SU$ with a symmetric flavor and fundamental flavors}
The spectrum which satisfies the anomaly matching conditions
$SU(F)^3$, $SU(F)^2 U(1)_X$, $SU(F)^2U(1)_B$, $SU(F)^2U(1)_R$,
$U(1)_X^3$, $U(1)_X$, $U(1)_B^3$, $U(1)_B$,  $U(1)_R^3$, $U(1)_R$,
$U(1)_X^2 U(1)_B$, $U(1)_X U(1)_B^2$, $U(1)_X^2 U(1)_R$, $U(1)_X U(1)_R^2$,
$U(1)_B^2U(1)_R$, $U(1)_B U(1)_R^2$, $U(1)_X U(1)_B U(1)_R$, \break
$SU(F)^2 Z_{2(k+1)F}$, $Z_{2(k+1)F}$, $Z_{2(k+1)F}^3$,
$U(1)_X^2 Z_{2(k+1)F}$, $U(1)_X Z_{2(k+1)F}^2$,   
$U(1)_B^2 Z_{2(k+1)F}$, \break $U(1)_B Z_{2(k+1)F}^2$,
$U(1)_R^2 Z_{2(k+1)F}$, $U(1)_R Z_{2(k+1)F}^2$,
$U(1)_X U(1)_B Z_{2(k+1)F}$, $U(1)_X U(1)_R Z_{2(k+1)F}$ and
$U(1)_BU(1)_R Z_{2(k+1)F}$ is given by
\begin{equation} 
\begin{array}{c|c|cccccc}
& SU(N) & SU(F) & SU(F) & U(1)_X & U(1)_B & U(1)_R & Z_{2F(k+1)}\\ \hline
X & \Ysymm & 1 & 1 & 1 & 0 & \frac{1}{k+1} & F \\
\bar{X} & \overline{\Ysymm} & 1 & 1 & -1 & 0 & \frac{1}{k+1} & F \\
Q & \Yfund & \Yfund & 1 & 0 & 1 & {\scriptstyle 1-\frac{F(1+2k)+2k-1}{F(1+k)}} & -(N+2)\\
\bar{Q} & \overline{\Yfund} & 1 & \Yfund & 0 & -1 &  
{\scriptstyle 1-\frac{F(1+2k)+2k-1}{F(1+k)}} & -(N+2) \\ \hline \hline
M_j & & \Yfund & \Yfund & 0 & 0 & 
\frac{2+2Fj-4k-2Fk}{F(1+k)}& {\scriptstyle 2jF-2(N+2)} \\
P_r & & \Ysymm & 1 & -1 & 2 & 
\frac{2+F-4k-2Fk+2Fr}{F(1+k)} &  {\scriptstyle (2r+1)F-2(N+2)}\\
\bar{P}_r & & 1 & \Ysymm & 1 & -2 & 
\frac{2+F-4k-2Fk+2Fr}{F(1+k)} &  {\scriptstyle (2r+1)F-2(N+2)} \\
B & & \overline{\Yfund} & 1 & \frac{N-F+1}{2} & 
F-1 & \frac{-1+F+2k+Fk}{F(1+k)} &  {\scriptstyle 
 1+F^2(1+k)+4k} \\
\bar{B} & & 1 & 
\overline{\Yfund} & -\frac{N-F+1}{2} & 
1-F & \frac{-1+F+2k+Fk}{F(1+k)} &   {\scriptstyle 
1+F^2(1+k)+4k} \\
T_i & & 1 & 1 & 0 & 0 & \frac{2i}{k+1} & 2iF \\
b  & & 1 & 1 & N-F & 2F & 
\frac{1}{1+k}&  {\scriptstyle -F(N+F+4)} \\
\bar{b} & & 1 & 1 & F-N & -2F & \frac{1}{1+k} &   
{\scriptstyle -F(N+F+4)}\\
\end{array},
\end{equation}
where  $j=0,\cdots ,k$, $r=0,\cdots ,k-1$, $i=1,\cdots ,k$, $N=(2k+1)F+4k-1$,
and the tree-level superpotential of the $SU(N)$ theory is 
${\rm Tr} (X\bar{X})^{k+1}$. The composite operators are given by
\begin{eqnarray}
&& M_j=Q(X\bar{X})^j\bar{Q}, \nonumber \\
&& P_r= Q(X\bar{X})^r\bar{X}Q,\nonumber \\
&& \bar{P}_r=\bar{Q}(X\bar{X})^rX\bar{Q},\nonumber \\
&& B= W_{\alpha}^{2k} Q^{-1+F+F k }\bar{Q}^{ F k} X^{k (F+k+F k)}\bar{X}^{k 
(-2+k+F k)},
\nonumber \\
&& \bar{B}=  W_{\alpha}^{2k} \bar{Q}^{-1+F+F k }{Q}^{ F k} \bar{X}^{k (F+k+
F k)} X^{k (-2+k+F k)}, \nonumber \\
&& T_i=(X\bar{X})^i,\nonumber \\
&& b =Q^{2F}X^{N-F},\nonumber \\
&& \bar{b} =\bar{Q}^{2F}\bar{X}^{N-F}.
\end{eqnarray}
The color indices $\kappa ,\lambda$ and flavor indices $i,j$ in $b$ are 
contracted by two epsilon tensors each: $b = \epsilon_{\kappa_{1} 
\cdots \kappa_{N}}\epsilon_{\lambda_{1} \cdots \lambda_{N}} \epsilon^{i_{1} 
\cdots i_{F}}\epsilon^{j_{1} \cdots j_{F}}Q^{\kappa_{1}}_{i_{1}} 
\cdots Q^{\kappa_{F}}_{i_{F}}Q^{\lambda_{1}}_{j_{1}} 
\cdots Q^{\lambda_{F}}_{j_{F}}X^{\kappa_{F+1}\lambda_{F+1}} \cdots 
X^{\kappa_{N}\lambda_{N}}$.
The gauge contraction in the operator $B$ is presumably given by
\begin{eqnarray*} B&=& (XW_{\alpha})^2 (X(X\bar{X})W_{\alpha})^2\cdots 
(X(X\bar{X})^{k-1}W_{\alpha})^2 Q^F ((X\bar{X})Q)^{F} \cdots
((X\bar{X})^kQ)^{F-1} \nonumber \\
&& (X\bar{Q})^F (X(X\bar{X})\bar{Q})^F \cdots  
(X(X\bar{X})^{k-1}\bar{Q})^F,\end{eqnarray*}
with one epsilon tensor
and similarly for the operator $\bar{B}$ with $Q \leftrightarrow \bar{Q}$ 
and $X\leftrightarrow \bar{X}$. For $k=0$ this theory again reproduces 
the s-confining $SU(F-1)$ theory with $F$ flavors. Note that in the
$k=0$ case there are no $P_i$ and $T_i$ operators, $B$ and $\bar{B}$ are just 
the  usual baryons $Q^F$ and $\bar{Q}^F$, and the operators 
$b$ and $\bar{b}$ are complete singlets under all symmetries except the 
$U(1)_R$ under which they carry $R$ charge one for $k=0$. Thus presumably
there is a mass term $b\bar{b}$  in the confining superpotential for $k=0$,
which eliminates these fields from the low-energy spectrum. It would be 
very interesting to see this explicitly happening by examining the
actual form of the confining superpotential for arbitrary values of $k$.

In addition to the above operators which describe the quantum moduli 
space, there is a classical flat direction
\begin{equation}
	X = v\left( \begin{array}{cccc} 1&&& \\ &1&& \\ &&\ddots& \\ &&&1
		\end{array} \right), \qquad
	\bar{X} = 0.
\end{equation}
This direction satisfies both the $F$-flatness $\bar{X}(X\bar{X})^{k} 
=(X\bar{X})^{k}X =0$ and the $D$-flatness $X^{\dagger} X \propto 1$ 
conditions, and corresponds to the operator ${\rm det}X$.  This flat 
direction, however, is removed from the moduli space quantum 
mechanically.  It leaves an $SO(N)$ gauge group unbroken, with $2F$ 
vectors and a symmetric tensor $\bar{X}$ with the superpotential ${\rm 
Tr} \bar{X}^{k+1}$.  This is precisely the particle content of the 
model discussed in Section~\ref{subsec:SOS}, except for the trace 
part of $\bar{X}$.  The gauge group, however, is larger, 
$N=(2k+1)F+4k-1 = [k(2F+4)-1]+F$.  Therefore the $SO(N)$ dynamics is 
expected to produce an Affleck--Dine--Seiberg type superpotential and 
the flat direction is lifted quantum mechanically for 
$F>1$.\footnote{The case $F=1$ is not fully understood.  From 
the analogy to the $SO(N)$ theory with $N-4$ vectors \cite{SO}, 
the low-energy 
$SO(6k)$ dynamics is expected to give two branches, one confining 
with spontaneous discrete symmetry breakdown \cite{discrete}
and no confining superpotential, and the other with 
run-away behavior.  Unlike in the previous Section, the ${\rm det}X$ and 
${\rm det}\bar{X}$ operators cannot be added to the spectrum without 
spoiling the anomaly matching conditions.  A likely possibility is 
that low-energy $SO(6k)$ dynamics is forced to choose the branch with 
the run-away behavior and the flat direction is removed from the 
quantum moduli space.  Another possibility is that the $Z_{2}$ 
instanton effect in $SU(6k)/SO(6k)$ \cite{CM} induces a term in the 
superpotential which leads to a run-away behavior.  We leave this issue 
for future investigation.}

\section{Conclusions\label{sec:concl}}

In this paper we have examined $N=1$ supersymmetric gauge theories which 
become confining after a suitable tree-level superpotential is added. 
These theories are obtained by examining the cases when the dual gauge groups
of Ref.~\cite{K,KS,KSS,Intr,LS,ILS} become trivial. We find that in all 
cases when the dual gauge group reduces to the trivial group the theory is
confining at the origin of the moduli space with a set of composites 
satisfying the 't Hooft anomaly matching conditions. A confining 
superpotential for these composites, 
which is necessary in order to reproduce the classical constraints,
is generated by the strong dynamics. This
superpotential can be fixed in most cases
by considering ``dressed quarks'', and examine the
classical constraints of the s-confining theory of these dressed flavors.
We have shown several examples of such theories, and in some cases we have 
completely fixed the confining superpotential, and showed how this 
superpotential is generated in the dual gauge group by instantons.

An interesting question is how to classify the sort of confining theories
examined in this paper. Most confining theories can be simply found,
because the matter content obeys an index condition 
$\mu_{matter}=\mu_{adjoint}+2$ for s-confining theories and 
$\mu_{matter}=\mu_{adjoint}$ for theories with a quantum modified 
constraint~\cite{s-conf}, where $\mu$ is the Dynkin index. 
However, in the theories presented in this paper the size of the 
confining gauge group  also depends on the form of the tree-level 
superpotential, therefore a simple index constraint does not seem to be
possible. This might be an advantage to these models compared to the 
ordinary s-confining ones, since the index constraint restricted the
possible s-confining theories to a rather small set, with limited sizes and 
varieties of global symmetries.
It would be very interesting to find a general way of 
analyzing these new confining theories without having to refer to the
dualities of the theories with a bigger matter content, and to establish which
confining spectra could be obtained this way.

\section*{Acknowledgements}

We thank Martin Schmaltz, Witold Skiba and John Terning for useful discussions.
C.C. thanks the Aspen Center for Physics for its hospitality where part of
this work was completed. C.C. is a research fellow of the Miller 
Institute for Basic Research in Science. H.M. is supported in part 
by the Alfred P. Sloan Foundation. This work is supported in part
the U.S. Department of Energy under Contract DE-AC03-76SF00098, and in part 
by the National Science Foundation under grant PHY-95-14797.

\end{document}